LA-UR-94-3641

# Symplectic Computation of Lyapunov Exponents

**Salman Habib[†] and Robert D. Ryne[⋆]**

[†] *T-6, Theoretical Astrophysics*
*and*
*T-8, Elementary Particles and Field Theory*
*Theoretical Division*
*Los Alamos National Laboratory*
*Los Alamos, NM 87545*

[⋆] *AOT-1, Accelerator Physics and Special Projects*
*Accelerator Operations and Technology Division*
*Los Alamos National Laboratory*
*Los Alamos, NM 87545*

## Abstract

A recently developed method for the calculation of Lyapunov exponents of dynamical systems is described. The method is applicable whenever the linearized dynamics is Hamiltonian. By utilizing the exponential representation of symplectic matrices, this approach avoids the renormalization and reorthogonalization procedures necessary in usual techniques. It is also easily extendible to damped systems. The method is illustrated by considering two examples of physical interest: a model system that describes the beam halo in charged particle beams and the driven van der Pol oscillator.

e-mail:
habib@predator.lanl.gov
ryne@lanl.gov

# 1 Introduction

Chaotic dynamical systems exhibit exponential divergence of initially nearby trajectories. This divergence is quantified by the Lyapunov exponents of the system which are obtained from linearizing the dynamics around a fiducial trajectory [1]. Over the past two decades or so there has been "intense activity" [2] directed toward the computation of these exponents resulting in several different numerical approaches [1][3][4]. The two obvious difficulties associated with the computation of Lyapunov exponents are: (1) exponential growth of the separation vector (between the fiducial and nearby trajectories) and (2) the exponential collapse of initially orthogonal separation vectors onto the direction of maximal growth. Most conventional methods overcome these hurdles by intermittent numerical rescaling and reorthogonalization (through, *e.g.*, the Gramm-Schmidt procedure [3]). Many chaotic systems are Hamiltonian or they can be transformed into a Hamiltonian system by suitable manipulations. However, none of the above general methods are designed to take advantage of this fact.

As is well known, the dynamics of classical Hamiltonian systems has an underlying symplectic structure [5]. In recent years symplectic methods have been applied with great success to classical dynamical problems. The field of accelerator dynamics has been revolutionized by the introduction of nonlinear symplectic maps as represented by Lie transformations [6][7]. Very long time integration of charged particle and planetary systems has been aided by the development of high order symplectic integration algorithms [8]. Recently a symplectic map-based approach for the calculation of Lyapunov exponents has been developed [9]. As shown below, this approach obviates *analytically* the need for rescaling and reorthogonalization in the numerical computation of the exponents.

# 2 The Method

Consider a 2-$m$ dimensional continuous-time dynamical system governed by the equations

$$\frac{d\mathbf{z}}{dt} = \mathbf{F}(\mathbf{z}, t), \tag{1}$$

where $\mathbf{z} = (z_1, z_2, \cdots, z_{2m})$ and similarly for $\mathbf{F}$. Let $\mathbf{z}_0$ denote some given fiducial trajectory. Define deviations from this trajectory by letting $\mathbf{Z} = \mathbf{z} - \mathbf{z}_0$, and linearize the above equations. The new set of equations for the deviation variables is

$$\frac{d\mathbf{Z}}{dt} = \mathbf{DF}(\mathbf{z}_0, t) \cdot \mathbf{Z}. \tag{2}$$

The approach described below can be used whenever this linearized set of equations is derivable from a Hamiltonian. From now on suppose that this is the case, and that one can write $\mathbf{Z} = (q_1, q_2, \cdots, q_m, p_1, p_2, \cdots, p_m)$, where $q_i$ and $p_i$ denote canonically conjugate coordinates and momenta, respectively. It follows that

$$\frac{d\mathbf{Z}}{dt} = -\{H, \mathbf{Z}\}, \tag{3}$$

where $\{,\}$ denotes the Poisson bracket, and where $H$ is a homogeneous quadratic polynomial in the $q_i$ and $p_i$. A system such as this is governed by a symplectic matrix $M$ that maps the initial variables $\mathbf{Z}^{in}$ into time-evolved variables $\mathbf{Z}(t)$,

$$\mathbf{Z}(t) = M(t)\mathbf{Z}^{in}. \tag{4}$$

Let $\Lambda$ be given by

$$\Lambda = \lim_{t\to\infty} \left(M\tilde{M}\right)^{1/2t}, \tag{5}$$

where $\tilde{M}$ denotes the matrix transpose of $M$. The Lyapunov exponents then equal the logarithm of the eigenvalues of $\Lambda$ [1].

It is easy to show that $M$ satisfies the equation of motion (See, for example, Ref. [11])

$$\frac{dM}{dt} = JSM, \tag{6}$$

where $S$ denotes the symmetric matrix given by

$$H(\mathbf{Z}, t) = \frac{1}{2} \sum_{i,j=1}^{2m} S_{ij} Z_i Z_j, \tag{7}$$

and where

$$J = \begin{pmatrix} 0 & \mathbf{1} \\ -\mathbf{1} & 0 \end{pmatrix}. \tag{8}$$

Here $\mathbf{1}$ denotes the $m \times m$ identity matrix. It follows that the evolution of $M\tilde{M}$ is governed by the equation

$$\frac{d}{dt} M\tilde{M} = JSM\tilde{M} - M\tilde{M}SJ. \tag{9}$$

Standard methods for obtaining the Lyapunov exponents deal with $M$, which is not real symmetric (hence the need for reorthogonalization) and which has exponentially growing elements. To avoid these difficulties one exploits the fact that $M$ is symplectic by making use of the exponential representation of symplectic matrices [6]: Any symplectic matrix $M$ can be written in the form

$$M = \mathrm{e}^{JS_a} \mathrm{e}^{JS_c} \tag{10}$$

where $S_a$ is a symmetric matrix that anticommutes with $J$ and $S_c$ is another symmetric matrix that commutes with $J$. It is important to note that the second matrix on the right hand side of (10) is in fact unitary, so that

$$M\tilde{M} = \mathrm{e}^{2JS_a}. \tag{11}$$

Note that this matrix has fewer degrees of freedom than $M$ and its eigenvectors are orthogonal. Rather than attempting to directly integrate Eqn. (9) which would still have a large numbers problem, focus attention instead on the exponent $JS_a$ in Eqn. (10). It is clear that now there is no large numbers problem since $JS_a$ already appears as an exponent.

To proceed further, one obvious approach is to use an explicit representation of $\exp(JS_a)$. Such a representation is well known for $Sp(2)$ and has recently been found for $Sp(4)$ (see the Appendix). Generalizations to $Sp(2m)$ are in progress [10]. To illustrate the method it is convenient to restrict attention to systems with a two-dimensional phase space. When driven, these represent the simplest continuous-time systems that can exhibit chaos. The most general two-dimensional symplectic matrix can be written in the form

$$\begin{aligned} M &= \mathrm{e}^{JS_a}\mathrm{e}^{JS_c} \\ &= \mathrm{e}^{\mu(B_2\cos a + B_3 \sin a)}\mathrm{e}^{bB_1}, \end{aligned} \tag{12}$$

where $a$, $b$ and $\mu$ are real coefficients and where the $B_i$ are basis elements of the Lie algebra $sp(2)$ [6]:

$$B_1 = \begin{pmatrix} 0 & 1 \\ -1 & 0 \end{pmatrix}, \quad B_2 = \begin{pmatrix} 0 & 1 \\ 1 & 0 \end{pmatrix}, \quad B_3 = \begin{pmatrix} 1 & 0 \\ 0 & -1 \end{pmatrix}. \tag{13}$$

It follows that

$$M\tilde{M} = \mathrm{e}^{2\mu(B_2\cos a + B_3\sin a)}. \tag{14}$$

Thus, one obtains,

$$\Lambda = \lim_{t\to\infty} \mathrm{e}^{(\mu/t)(B_2\cos a + B_3 \sin a)}. \tag{15}$$

Finally, it is easily shown that the eigenvalues of this matrix are $\mathrm{e}^{\pm\mu/t}$. The Lyapunov exponents are then equal to $\pm\mu/t$ in the limit $t \to \infty$. With this convenient choice of variables, the explicit representation of $M\tilde{M}$ is given by

$$M\tilde{M} = \begin{pmatrix} \cosh 2\mu + \sin a \sinh 2\mu & \cos a \sinh 2\mu \\ \cos a \sinh 2\mu & \cosh 2\mu - \sin a \sinh 2\mu \end{pmatrix}. \tag{16}$$

The unknown quantities $a$ and $\mu$ can grow in time at most as $O(t)$. Differential equations for these quantities can be obtained by returning to Eqn. (9), the dynamical equation for $M\tilde{M}$.

For simplicity, assume that $H$ contains no term proportional to $qp$, so that the matrix $S$ in (7) is of the form

$$S = \begin{pmatrix} s_{11} & 0 \\ 0 & s_{22} \end{pmatrix}. \tag{17}$$

After some manipulation, Eqns. (7)–(9) lead to the following:

$$\begin{aligned} \frac{d\mu}{dt} &= \frac{1}{2}(s_{22} - s_{11})\cos a, \\ \frac{da}{dt} &= s_{11} + s_{22} - (s_{22} - s_{11})\sin a \coth \mu. \end{aligned} \tag{18}$$

From the initial condition $M(0) = I$, if one chooses $\mu(0) = 0$, then $\cos^2 a(0) = 1$, *i.e.*, $a(0) = 0$ or $\pi$. These differential equations form the basis of the method for calculating the Lyapunov exponents of Hamiltonian systems: They are stepped forward in time numerically till some desired convergence for the exponents, $\pm\mu/t$, is achieved. It will be shown later how to apply the method to certain non-Hamiltonian systems.

# 3 Applications: Two Examples

As a first concrete example, consider the newly developed "core-halo" model which describes beam halo in mismatched charged particle beams [12]. The transverse equation of motion for a halo particle in this model, assuming constant external focusing, is

$$\ddot{x} + x - (1 - \eta^2) f(x, r(t)) = 0 \tag{19}$$

where $x$ is the position variable for a halo particle, $f(x, r(t))$ is the force due to the space charge of the beam core, and $r(t)$ is the time dependent *rms* radius of the core. The core radius is assumed to follow the envelope equation

$$\ddot{r} + r - \frac{1 - \eta^2}{r} - \frac{\eta^2}{r^3} = 0. \tag{20}$$

Here units have been chosen so that the time independent solution of (20) (*i.e.*, a matched beam) is given by $r = 1$. In these units $\eta = 0$ corresponds to the space charge dominated regime, while $\eta = 1$ corresponds to the emittance dominated regime. Now assume

$$f(x, r) = \frac{x}{x^2 + r^2} \tag{21}$$

which has the correct asymptotic behavior: the force is linear when $x \ll r$, and it is inversely proportional to $x$ when $x \gg r$. The Eqns. (19) and (20) describe a driven nonlinear system with a mixed phase space as demonstrated by the stroboscopic plot shown in Fig. 1. The presence of a chaotic band is important because particles initially in the core can leak through the broken separatrix and be carried to large amplitudes. The presence of such large amplitude particles can cause unacceptably high radioactivation levels in high intensity linacs planned for future accelerator-driven technologies [13]. Leakage through the separatrix can be enhanced through particle collisions and recent work has shown that this rate is controlled by the Lyapunov exponent [14]. The Lyapunov exponent for this system may be computed by integrating (18) with

$$
\begin{aligned}
s_{11} &= 1-\left(1-\eta^2\right)\left(\frac{1}{x_0^2+r^2}-\frac{2x_0^2}{\left(x_0^2+r^2\right)^2}\right) \\
s_{22} &= 1
\end{aligned} \tag{22}
$$

where $x_0$ denotes the fiducial trajectory. Fig. 2 displays the result for the Lyapunov exponent against time. The slow convergence of the exponent to its asymptotic value is typical of Hamiltonian systems.

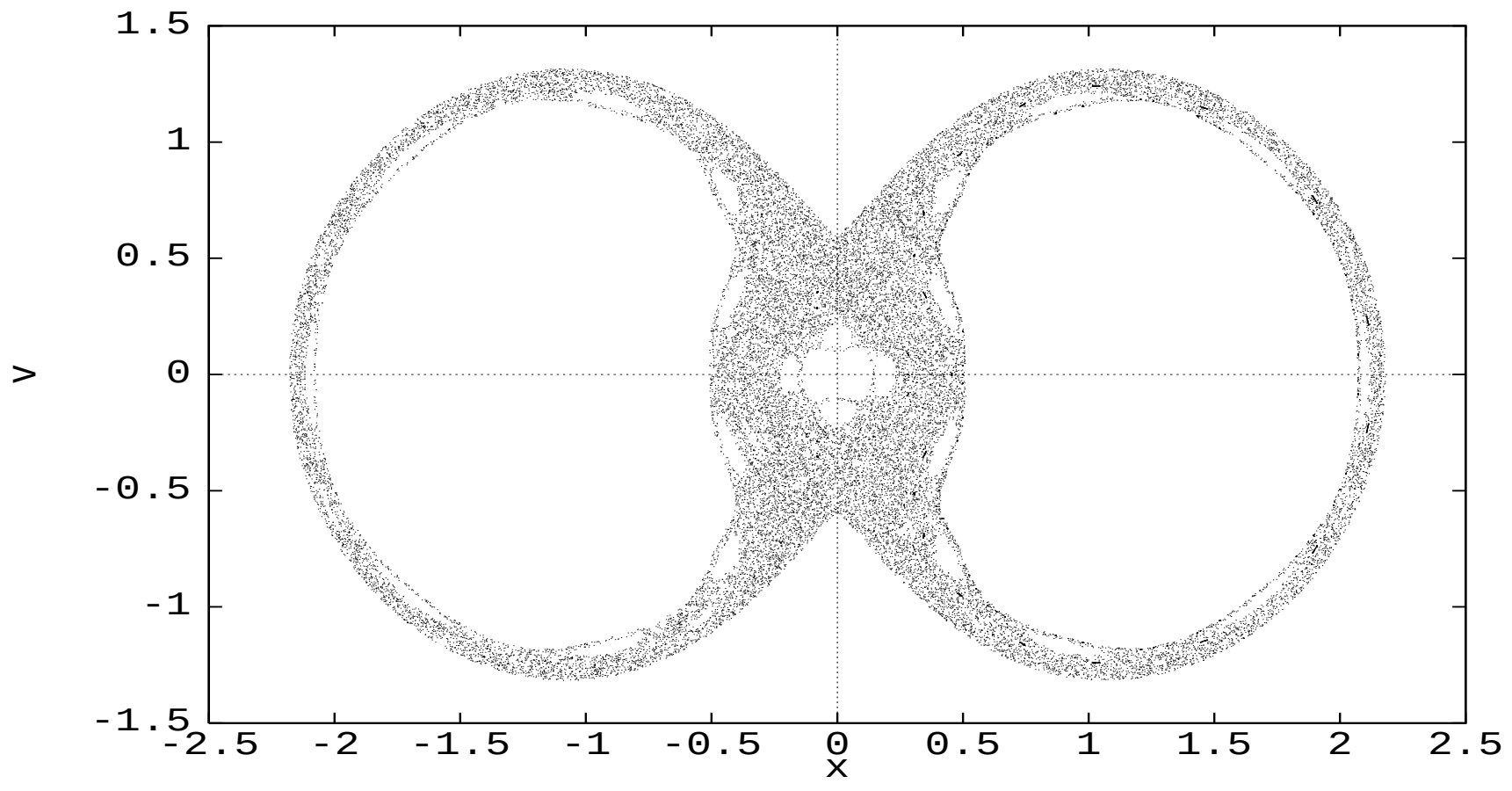

Figure 1: Stroboscopic plot of the chaotic sea in the core-halo model. Snapshots were taken at successive beam minima for 32 test particles. Parameter values were $r(0) = 0.6$, $\dot{r}(0) = 0$, and $\eta = 0.2$.

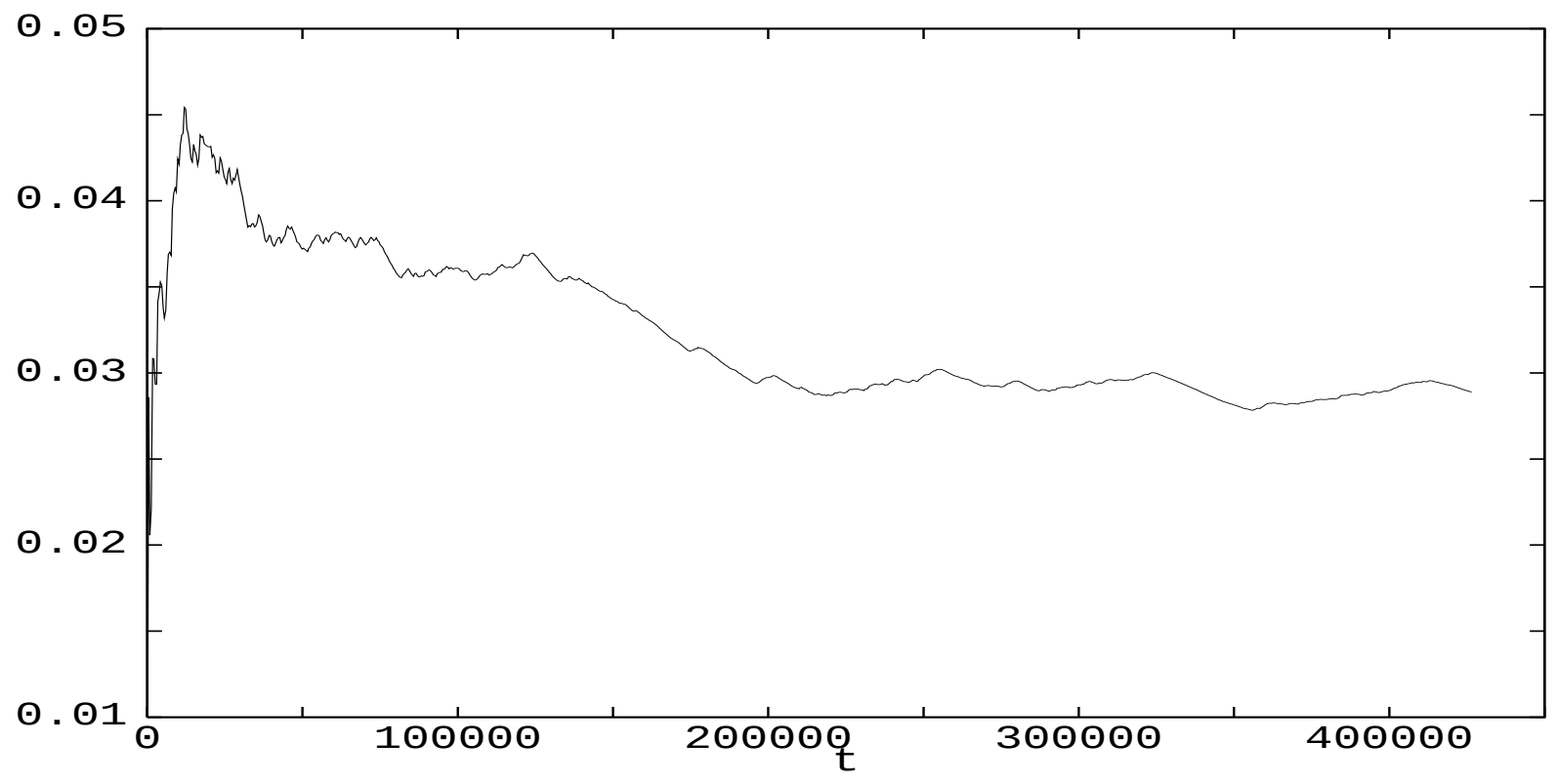

Figure 2: Positive Lyapunov exponent for the core-halo model in a typical run. Parameters are the same as in Fig. 1. The simulation was run for $10^5$ periods of the driving force using 100 integration steps per period for Eqns. (18) with a third-order Runge-Kutta algorithm. The envelope equation and the fiducial trajectory were integrated with a fourth-order symplectic algorithm using 200 steps per period.

So far only explicitly Hamiltonian systems were considered. However, the only real requirement for using the method is that the linearized deviation equations in *some* variables be Hamiltonian. This allows for the inclusion of damped systems in the scheme. As an example, consider the following general driven nonlinear oscillator

$$\ddot{x}+\lambda\left(1-\epsilon x^2\right)\dot{x}+V'(x)=a\cos(\omega t). \tag{23}$$

By appropriate choices of $\lambda$, $\epsilon$, and $V(x)$, this reduces to an assortment of well-known equations including van der Pol ($\lambda<0$, $\epsilon=1$, $V(x)=(1/2)x^2$), Duffing ($\lambda>0$, $\epsilon=0$, $V(x)=\alpha x^2+\beta x^4$), and the damped driven pendulum ($\lambda>0$, $\epsilon=0$, $V(x)=1-\cos(x)$). In terms of the deviation variable $\delta$, the linearization of (23) yields

$$\ddot{\delta}+\lambda\left(1-\epsilon x_0^2\right)\dot{\delta}+\left(V''(x_0)-2\epsilon\lambda x_0\dot{x}_0\right)\delta=0 \tag{24}$$

where $x_0$ represents the fiducial trajectory. Introducing the new variable $\Delta$ defined through

$$\Delta=\delta \mathrm{e}^{-g(t)} \tag{25}$$

where

$$\dot{g}=-\frac{1}{2}\lambda\left(1-\epsilon x_0^2\right), \tag{26}$$

Eqn. (24) reduces to that describing an undamped oscillator with time dependent frequency,

$$\ddot{\Delta}+\left(V''(x_0)-\epsilon\lambda x_0\dot{x}_0-\frac{1}{4}\lambda^2\left(1-\epsilon x_0^2\right)^2\right)\Delta=0. \tag{27}$$

It is now straightforward to proceed in the usual way: for linear damping ($\epsilon = 0$) the Lyapunov exponents $\chi_\pm$ of this system are given by

$$\chi_\pm = -\frac{1}{2}\lambda \pm \lim_{t\to\infty} \frac{1}{t}\mu_0(t) \tag{28}$$

where $\mu_0$ follows from solving (18) for the system defined by (27). When $\epsilon \neq 0$,

$$\chi_\pm = \lim_{t\to\infty} \frac{1}{t}\left(g(t) \pm \mu_0(t)\right) \tag{29}$$

modulo terms that are exponentially suppressed at late times.

Figs. 3(a) and 3(b) show the Lyapunov exponents of the van der Pol system. For the chosen set of parameters these results are in agreement with those of Ref. [4]. In contrast with the results shown in Fig. 2, the convergence of the exponents is much faster, as is typical of non-weakly damped systems.

# 4 Conclusion

To summarize, a method for computing Lyapunov exponents that exploits the underlying symplectic structure of Hamiltonian dynamics has been reviewed. Just as symplectic integrators are not a panacea for all time integration problems, this method does not have universal applicability and advantages. However, when applicable, the method has certain advantages over standard techniques, most importantly the lack of systematic errors associated with intermittent reorthogonalization and rescaling.

# 5 Acknowledgements

The authors acknowledge useful discussions with Alex Dragt, Michael Mattis, and Michael Nieto. This work was supported by the U. S. Department of Energy at Los Alamos National Laboratory and by the Air Force Office of Scientific Research. The authors also acknowledge use of the facilities provided by the Advanced Computing Laboratory at Los Alamos National Laboratory.

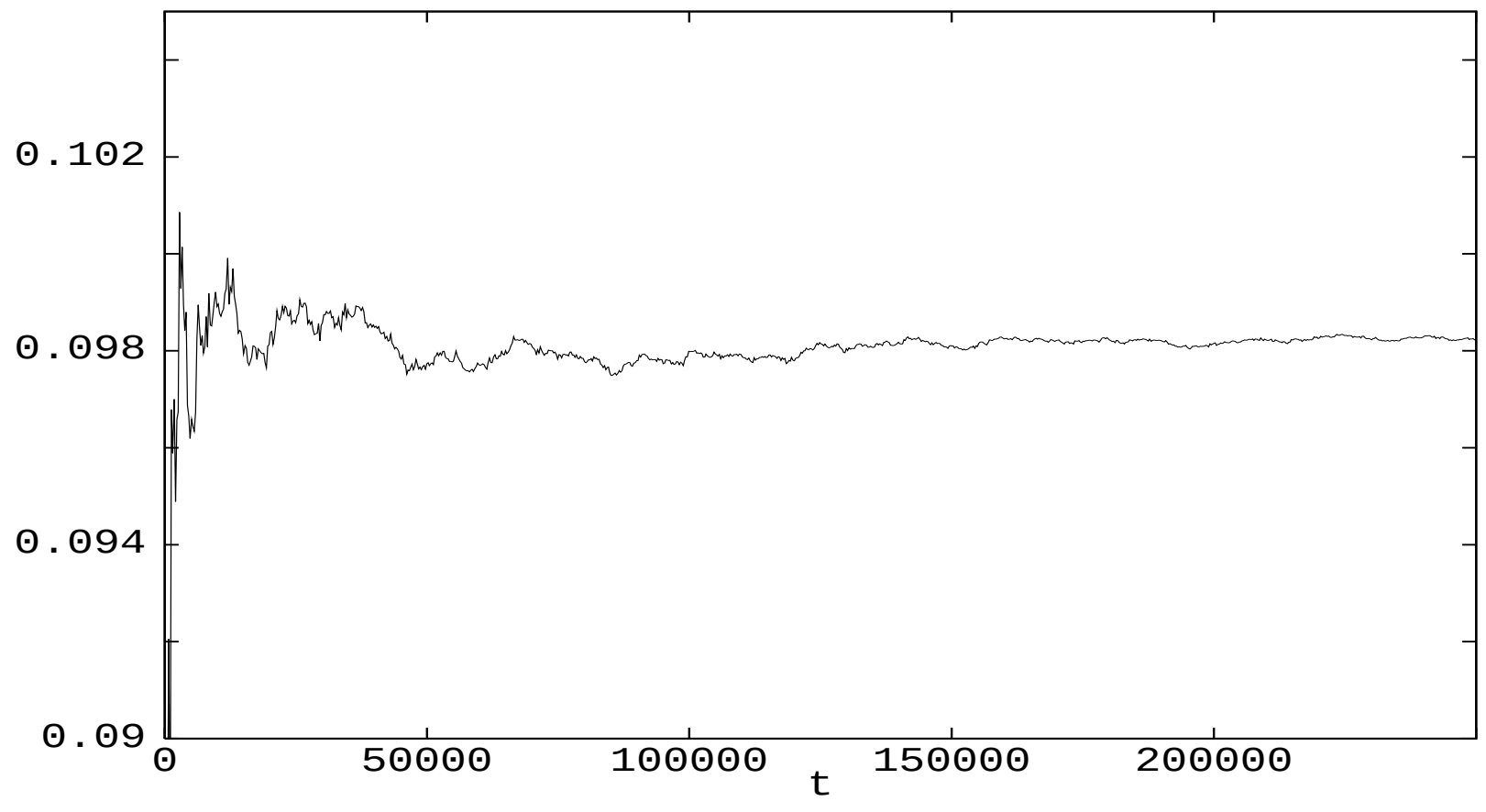

Figure 3: (a)Positive Lyapunov exponent for the van der Pol oscillator with parameters $\lambda = -5$, $a = 5$, and $\omega = 2.466$ (taken from Ref. [4]). The simulation was run for $10^5$ periods of the driving force using 100 million time steps for Eqns. (18) and 200 million time steps for the fiducial trajectory. The integrators were third-order Runge-Kutta and fourth-order symplectic, respectively.

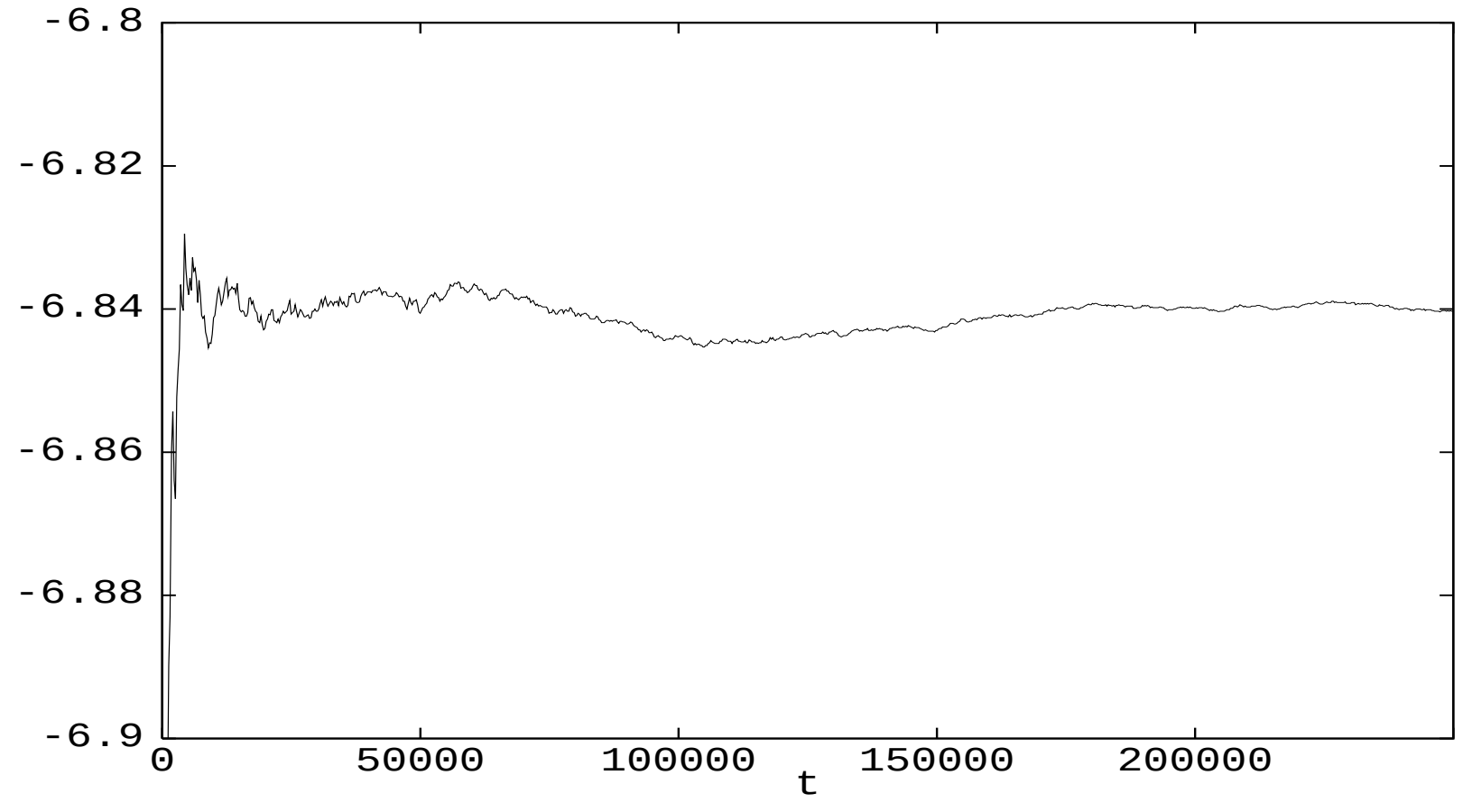

Figure 3: (b)Negative Lyapunov exponent for the van der Pol oscillator with the same parameters as in Fig. 3(a).

## A Appendix: Summary of Sp(4) Results

There exists a particularly convenient way of organizing the basis elements of $sp(4)$ in terms of a triplet of matrices (the $F_i$, $G_i$, and the $B_i$) each triplet consisting in turn of three matrices. The $B_i$ belong to the unitary sector which is irrelevant to the computation of the Lyapunov exponents. The most general $(4\times 4)$ symplectic matrix then turns out to be of the form $\exp(\mathbf{a}\cdot\mathbf{F}+\mathbf{b}\cdot\mathbf{G})$ times a unitary matrix, where

$$\begin{aligned}\mathbf{a} &= \{a_1,a_2,a_3\}, \qquad \mathbf{b}=\{b_1,b_2,b_3\}\\ \mathbf{F} &= \{F_1,F_2,F_3\}, \qquad \mathbf{G}=\{G_1,G_2,G_3\}.\end{aligned} \tag{30}$$

Here, the $a_i$ and the $b_i$ are scalars, and the explicit forms of the matrices $F_i$ and $G_i$ are given below.

It turns out to be possible to resum the formal exponential above and to obtain the following result [10]:

$$\begin{aligned}M &\equiv \exp(\mathbf{a}\cdot\mathbf{F}+\mathbf{b}\cdot\mathbf{G})\\ &= \frac{1}{2}(\cosh[u_x]+\cosh[u_y])\mathbf{I}\\ &\quad +\frac{1}{2}(\mathrm{sch}[u_x]+\mathrm{sch}[u_y])\,(\mathbf{a}\cdot\mathbf{F}+\mathbf{b}\cdot\mathbf{G})\\ &\quad +\frac{4}{(u_x^2-u_y^2)}(\cosh[u_x]-\cosh[u_y])\,(\mathbf{a}\times\mathbf{b})\cdot\mathbf{K}\\ &\quad +\frac{4}{(u_x^2-u_y^2)}(\mathrm{sch}[u_x]-\mathrm{sch}[u_y])\,([\mathbf{b}\times(\mathbf{a}\times\mathbf{b})]\cdot\mathbf{F}+[\mathbf{a}\times(\mathbf{b}\times\mathbf{a})]\cdot\mathbf{G})\end{aligned} \tag{31}$$

where "$\cdot$" represents the usual vector dot product, "$\times$" represents the vector cross product, $\mathbf{I}$ is the $(4\times 4)$ identity matrix and

$$\begin{aligned}\mathrm{sch}[x] &\equiv \sinh[x]/x,\\ u_x &= \sqrt{2s^2+4v},\\ u_y &= \sqrt{2s^2-4v},\\ s^2 &= \mathbf{a}\cdot\mathbf{a}+\mathbf{b}\cdot\mathbf{b},\\ v^2 &= (\mathbf{a}\times\mathbf{b})\cdot(\mathbf{a}\times\mathbf{b}).\end{aligned} \tag{32}$$

A new triplet of matrices, the $K_i$ appears. This is because the $F_i$ and the $G_i$ do not form a closed subalgebra. The $K_i$ are idempotent and unitary. Note that all the $F_i$, $G_i$ and $K_i$ are traceless. Thus, quite trivially, $Tr(M)=2(\cosh[u_x]+\cosh[u_y])$ and the eigenvalues of $M$ are $\exp[u_x], \exp[-u_x], \exp[u_y], \exp[-u_y]$.

Finally, the explicit forms of the above matrices are given by:

$$K_1 = \begin{pmatrix} -1 & 0 & 0 & 0 \\ 0 & 1 & 0 & 0 \\ 0 & 0 & -1 & 0 \\ 0 & 0 & 0 & 1 \end{pmatrix}, K_2 = \begin{pmatrix} 0 & -1 & 0 & 0 \\ -1 & 0 & 0 & 0 \\ 0 & 0 & 0 & -1 \\ 0 & 0 & -1 & 0 \end{pmatrix}, K_3 = \begin{pmatrix} 0 & 0 & 0 & -1 \\ 0 & 0 & 1 & 0 \\ 0 & 1 & 0 & 0 \\ -1 & 0 & 0 & 0 \end{pmatrix} \tag{33}$$

$$F_1 = \begin{pmatrix} 0 & -1 & 0 & -1 \\ -1 & 0 & -1 & 0 \\ 0 & -1 & 0 & 1 \\ -1 & 0 & 1 & 0 \end{pmatrix}, F_2 = \begin{pmatrix} 1 & 0 & 1 & 0 \\ 0 & -1 & 0 & -1 \\ 1 & 0 & -1 & 0 \\ 0 & -1 & 0 & 1 \end{pmatrix}, F_3 = \begin{pmatrix} 1 & 0 & -1 & 0 \\ 0 & 1 & 0 & -1 \\ -1 & 0 & -1 & 0 \\ 0 & -1 & 0 & -1 \end{pmatrix} \tag{34}$$

$$G_1 = \begin{pmatrix} 0 & -1 & 0 & 1 \\ -1 & 0 & 1 & 0 \\ 0 & 1 & 0 & 1 \\ 1 & 0 & 1 & 0 \end{pmatrix}, G_2 = \begin{pmatrix} 1 & 0 & -1 & 0 \\ 0 & -1 & 0 & 1 \\ -1 & 0 & -1 & 0 \\ 0 & 1 & 0 & 1 \end{pmatrix}, G_3 = \begin{pmatrix} -1 & 0 & -1 & 0 \\ 0 & -1 & 0 & -1 \\ -1 & 0 & 1 & 0 \\ 0 & -1 & 0 & 1 \end{pmatrix} \tag{35}$$